\documentclass[conference]{IEEEtran}

\usepackage[utf8]{inputenc}

\usepackage{xspace}
\usepackage{amsfonts,amsmath,amssymb} 
\usepackage{times}
\usepackage{algorithm}
\usepackage{algpseudocode}
\usepackage{algorithmicx}
\usepackage[usenames,dvipsnames]{color}

\usepackage{mymacros}

\usepackage[defblank]{paralist} 

\usepackage{graphicx}
\usepackage{subfigure}

\usepackage{mathptmx} 

\newcommand{\InvCounter}{\emph{Bounded Counter}}
\newcommand{\InvCounters}{\emph{Bounded Counters}}
\newcommand{\InvCounterCRDT}{\emph{Bounded Counter}}

\newcommand{\comment}[1]{} 

\newcommand{\todo}[1]{} 

\renewcommand{\todo}[1]{\textbf{#1}} 

\newcommand{\url}[1]{\emph{#1}} 

\begin{document}

\author{\IEEEauthorblockN{Valter Balegas, Diogo Serra, Sérgio Duarte\\Carla Ferreira, Rodrigo Rodrigues, Nuno Preguiça}
\IEEEauthorblockA{NOVA LINCS/FCT/Universidade Nova de Lisboa}
\and
\IEEEauthorblockN{Marc Shapiro, Mahsa Najafzadeh}
\IEEEauthorblockA{INRIA / LIP6}}

\title{Extending Eventually Consistent Cloud Databases for Enforcing Numeric Invariants}
\maketitle

\begin{abstract}

Geo-replicated databases often operate under the
principle of eventual consistency to offer high-availability with low latency
on a simple key/value store abstraction. Recently, some have adopted 
commutative data types to provide seamless reconciliation for special purpose data types, such as counters. 
Despite this, the inability to enforce numeric invariants across all replicas
still remains a key shortcoming of relying on the limited guarantees of eventual consistency storage. 

We present a new replicated data type, called bounded counter,
which adds support for numeric invariants to eventually consistent geo-replicated databases.
We describe how this can be implemented on top of existing cloud stores without modifying them,
using Riak as an example. Our approach adapts ideas from escrow transactions to
devise a solution that is decentralized, fault-tolerant and fast. Our evaluation shows much lower 
latency and better scalability than the traditional approach of
using strong consistency to enforce numeric invariants, thus
alleviating the tension between consistency and availability.

\end{abstract}





\section{Introduction}	
\label{sec:introduction}

Scalable cloud databases with a key/value store interface 
have emerged as the platform of choice for providing online services that operate on a global scale,
such as Facebook~\cite{cassandra}, Amazon~\cite{dynamo}, or Yahoo!~\cite{pnuts}.
In this context, a common technique for improving the user experience is geo-replication~\cite{dynamo,pnuts,walter}, i.e., maintaining copies of application data and logic in multiple data centers scattered across the globe. This decreases the latency for handling user requests by routing them
to nearby data centers, but at the expense of resorting to weaker data consistency guarantees, in order to avoid a costly coordination across replicas
for executing operations.

When executing under such weaker consistency models, applications have to deal with concurrent operations executing without being aware of each other, which implies that a merge strategy is required for reconciling concurrent updates. 
A common approach is to rely on a \emph{last writer wins} strategy \cite{cops,eiger,cassandra}, but this is not appropriate in all situations. 
A prominent example is the proper handling of counters, which are a useful abstraction
for implementing features such as \emph{like} buttons, votes and ad and page views, and all sorts of resource counting. For counters, using \emph{last writer wins} leads to lost updates, breaking the intended semantics.
To address this limitation, cloud databases, such as Cassandra \cite{cassandra:counters}, DynamoDB and Riak\cite{riak:counters}, have extended their interfaces with support for correct counters, implemented 
using specific merge algorithms.


Even though these approaches provide a principled handling of concurrent updates to counter objects, they fall short on supporting the enforcement of crucial invariants or database integrity constraints, which are often required for maintaining correct operation~\cite{redblue}. 
Real world examples where enforcing invariants is essential are advertisement services,
virtual wallets or to maintain stocks.
However, enforcing this condition using counters implemented on eventually consistent 
cloud database is impossible. This is because counter updates can occur concurrently, 
making it impossible to detect if the limit is exceeded before the operation concludes.


Maintaining this type of invariants would be trivial in systems that offer strong consistency guarantees, namely those that serialize all updates, and therefore preclude that two operations execute without seeing the effects of one another \cite{spanner,walter,redblue}. 
The problem with these systems is that they require coordination among replicas, leading to an increased latency. In particular, in a geo-replicated scenario, this latency may amount to hundreds of milliseconds, which suffices to impact application usability~\cite{Schurman09latency}.

In this paper we show that it is possible to achieve the best of both worlds, i.e.,
that fast geo-replicated operations on counters can coexist with strong invariants.
To this end, we propose a novel abstract data type 
called a \InvCounterCRDT{}.
This replicated object, like conventional CRDTs~\cite{crdts}, allows for 
operations to execute locally, automatically merges concurrent updates, and, in contrast to previous CRDTs, also
enforces numeric invariants while avoiding coordination in most cases.
Implementing \InvCounterCRDT{} in a fast and portable way required overcoming
a series of challenges, which form the main technical contributions of this
work.

First, we propose an extension to the 
main idea behind escrow transactions~\cite{escrow}, which is to
partition the difference between the current value of a counter and the limit to be enforced among 
existing replicas. These parts are distributed among
replicas, who can locally execute operations that do not exceed their allocated
part.
Unlike previous solutions that include some central authority
and are often based on synchronous interactions between 
nodes \cite{escrow,demarcation,mobisnap,exo-leasing}, our approach is completely
decentralized and asynchronous, with each replica relying only on a local and
possibly stale view of the information and on peer-to-peer asynchronous interactions. 
This design makes it easy to
deploy our system, since we do not need to add a new master server
(or replica group) that
controls the allocation of operations on the counter.
Furthermore, this avoids situations where the temporary 
unreachability of the data center where the master server 
is located can prevent operations from making progress.

Second, and building on the fact that we did not have to add any new
master servers to enforce invariants, we show how it is possible to
layer our design on top of existing eventually consistent storage
systems, while making very few assumptions about the underlying
system. In particular, we only assume that the underlying storage
system executes operations in a serializable way in each replica (not
necessarily by the same order across replicas) and that it provides a
reconciliation mechanism for merging concurrent updates. This makes
our solution generic and portable, but raises the bar
for achieving a performance that is comparable to directly accessing
the underlying storage.
Furthermore, we propose two alternative designs, where the first one
is implemented using only a client-side library, whereas the second one includes
a server side component deployed in a distributed hash table, which provides 
better scalability by minimizing the number of operations executed 
in the underlying storage system.



The evaluation of our prototypes running on top of Riak shows that: 
\begin{inparaenum}
\item when compared to using weak consistency, our approach 
with the cache and a write batching mechanism has higher
throughput with a very small increase in latency, while 
guaranteeing that invariants are not broken;
\item when compared to using strong consistency, our approach can 
enforce invariants without paying the latency price for replica coordination,
which is considerable for all but the local clients;
\item the client based design performs well under low contention, 
but does not scale when contention on the same counter is large;
\item the server based design scales well with the number of clients 
executing operations, providing even higher throughput than 
weak consistency.
\end{inparaenum}









The remainder of the paper is organized as follows. 
Section~\ref{sec:model} overviews our solution and its requirements; 
Section~\ref{sec:crdt} introduces the \InvCounter{} CRDT; 
Section~\ref{sec:deploy:riak} presents our two designs that extend Riak with numeric invariant preservation; 
Section \ref{sec:eval} evaluates our prototypes; 
Section \ref{sec:extension} discusses extensions to the proposed design; 
Section \ref{sec:related} discusses related work; and 
Section \ref{sec:conclusions} concludes the paper.



\section{System Overview}\label{sec:model}


\subsection{Assumptions}

We target a typical geo-replicated scenario, with copies of application data and logic 
maintained in multiple data centers (DC) scattered across the globe.
End clients contact the closest DC for executing application operations in 
the application server running in that DC. 
The execution of this application logic leads to issuing a sequence of operations 
on the database system where application data resides. 

We consider that system processes
(or nodes) are connected by an asynchronous network (i.e., subject to arbitrary
delays, including partitions).
We assume a finite set $\Pi = {p_0, p_1, \ldots,p_{n-1}}$ of processes who may fail by crashing.
A crashed process may remain crashed forever, or may recover with its persistent memory intact.
A non-crashed process is said to be {\em correct}.

For simplicity, our presentation considers a single data object replicated in all 
processes of $\Pi$, with $r_i$ representing the replica of the object at process $p_i$.
The model trivially generalizes to the case where multiple data objects exist -- in 
such a case, for each object $o$, we need to consider only the set $\Pi^{o}$ of the processes that replicate $o$.

%
%
%
%

\subsection{System API}

\begin{figure}[t] 
{\footnotesize \centering
\begin{verbatim}
% Regular data operations
get(key): object | fail
put(key, object): ok | fail

% Bounded Counters operations
create(key, type, bound): ok | error
read(key): integer | error
inc(key, delta, flag): ok | fail | retry
dec(key, delta, flag): ok | fail | retry 
\end{verbatim}
\vspace{-2ex}
}
\caption{System API.}
\label{fig:API}
\vspace{-2ex}
\end{figure}

Our middleware system is built on top of a weakly-consistent key-value database. 
Figure \ref{fig:API} summarizes the programming interface of our system, with the
usual \emph{get} and \emph{put} operations for accessing regular data, and additional
operations for creating a new \InvCounter{}, reading its current state, and incrementing 
or decrementing its value.
As any other data, InvCounters{} are identified in all operations by an 
opaque key.

The \emph{create} operation creates a new bounded counter. 
The \emph{type} argument specifies if it is an \emph{upper-} or a \emph{lower-} \InvCounter{}, 
and the \emph{bound} argument provides the global invariant limit to be maintained -- e.g., 
\emph{create(``X'', upper, 1000)} creates a \InvCounter{} that maintains the invariant 
that the value must be \emph{smaller or equal to 1000}. 
The counter is initialized to the value of the bound.


The \emph{read} operation returns the current value of the given counter. 
The returned value is computed based on local information and it may not be globally accurate. 
To update a counter, the application submits \emph{inc} or \emph{dec} operations. 
These operations include a \emph{flag} to decide on whether the execution is strictly local or whether global execution is attempted.
In both cases, the operation attempts to run locally first. When the local information cannot guarantee that the value remains within bounds, in the case of a strictly local operation, the API returns an error and a hint regarding whether global execution is likely to succeed; otherwise, in the case of a global operation, the system tries to contact remote replicas to safely execute to operation and only returns an error if this coordination with remote replicas cannot ensure the preservation of the invariant (namely when the counter has reached its limit).



\subsection{Consistency Guarantees}

We build our middleware on top of an eventually consistent database, 
extending the underlying guarantees with invariant preservation for counters.
In particular, the eventual consistency model means that the outcome of each operation reflects
the effects of only a subset of the operations that all clients have previously invoked~--~these are the
operations that have already been executed by the replica that the client has contacted. However, for
each operation that successfully returns at a client, there is a point in time after which its effect
becomes visible to every operation that is invoked after that time, i.e., operations are eventually
executed by all replicas.

In terms of the invariant preservation guarantee, this means that the bounds on the counter
value are never violated, neither \emph{locally} nor \emph{globally}. 
By locally, this means that the bounds must
be obeyed in each replica at all times, i.e., the subset of operations seen by the replica must
obey:\\
\vspace{.2em} 
lower bound $\le$ initial value $ + \sum \textit{inc} - \sum  \textit{dec} \le$ upper bound.\\
\vspace{.2em} 
\noindent By globally, this means that, at any instant in the execution of the system, when
considering the union of all the operations executed by each replica, the same bounds must hold.


Note that the notion of causality is orthogonal to our design and guarantees, in the sense that if the underlying storage system offers causal consistency, then we  also provide numeric invariant-preserving causal consistency.

\subsection{Enforcing Numeric Invariants}\label{sec:model:overview}

To enforce numeric invariants, our design borrows ideas from 
the escrow transactional model~\cite{escrow}. 
The key idea of this model is to consider that the difference between the 
value of a counter and its bound can be seen as a set of rights to execute operations. 
For example, in a counter, $n$, with initial value $n = 40$ and invariant $n \geq 10$, 
there are $30$ ($40 - 10$) rights to execute decrement operations. 
Executing \emph{dec(5)} consumes $5$ of these rights. 
Executing \emph{inc(5)} creates $5$ rights.
In this model, these rights are split among the replicas of the 
counter -- e.g. if there are $3$ replicas, each replica can be assigned $10$ rights.
If the rights needed to execute some operation exist in the local replica, 
the operation can safely execute locally, knowing that the global 
invariant will not be broken -- in the previous example, 
if the decrements of each replica are less or equal to $10$, it follows that the total 
number of decrements does not exceed $30$, and therefore the invariant is preserved.
If not enough rights exist, then either the operation fails or additional rights 
must be obtained from other replicas.

Our approach encompasses two components that work together to achieve the goal of our system: 
a novel data structure, the \InvCounter{} CRDT, to maintain the necessary information for 
locally verifying whether it is safe to execute an operation or not; and 
a middleware layer to manipulate instances of this data structure stored in
the underlying cloud database. 
The first component is detailed in Section~\ref{sec:crdt}, 
while alternative designs to the second part are detailed in Section~\ref{sec:deploy:riak}.

\section{Design of Bounded Counter CRDT}\label{sec:crdt}

This section presents the design of \InvCounter{}, a CRDT that can be used to enforce numeric invariants without requiring coordination for most operation executions. Instead, coordination is normally executed outside of the normal execution flow of an operation and amortized over multiple operations.




\subsection{CRDT Basics}

Conflict-free replicated data types (CRDTs)~\cite{crdts} are a class of distributed data
types that allow replicas to be modified without coordination, while
guaranteeing that replicas converge to the same correct value after
all updates are propagated and executed in all replicas.  

Two types of CRDTs have been defined: \emph{operation-based CRDTs},
where modifications are propagated as operations (or patches) and executed on every
replica; and \emph{state-based CRDTs}, where modifications are
propagated as states, and merged with the state of every replica.

In this work, we have adopted the state-based model, as we have built our 
work on top of a key-value store that synchronizes replicas by propagating the 
state of the database objects. In this model,
one operation submitted in one site executes in the local replica.
Updates are propagated among replicas 
in peer-to-peer interactions, where a replica $r_1$ propagates its state 
to another replica $r_2$, which merges its local and received state, 
by executing the \emph{merge} function.

State-based CRDTs build on the definition of 
a join semi-lattice (or just semi-lattice), which is a
partial order $\SLleq$ equipped with a \emph{least upper bound} (LUB)
$\SLLUB$ for all pairs:
  $m = x \SLLUB y$ is a Least Upper Bound of
  $\{x,y\}$ under $\SLleq$ iff 
  $x \le m \land y \le m \wedge \forall m', 
     x \le m' \land y \le m' 
         \implies
       m \le m'$.

It has been proven that a sufficient condition for guaranteeing the convergence 
of the replicas of state-based CRDTs is that the object conforms the properties of 
a monotonic semi-lattice object \cite{crdts}, in which:
\begin{inparaenum}[\em (i)]
\item 
  The set $S$ of possible states forms a semi-lattice ordered by \SLleq;
\item
  The result of merging state $s$ with remote state $s'$ is the result of
  computing the LUB of the two
  states in the semi-lattice of states, i.e., $merge(s,s') = s \SLLUB s'$;
\item
  The state is monotonically non-decreasing across updates, i.e., for any update
  $u$, $s \SLleq u(s)$.
\end{inparaenum}

\subsection{Bounded Counter CRDT}

\begin{figure}[t]
\begin{algorithmic}
\PayloadML{integer[$n$][$n$] $R$, integer[$n$] $U$, integer $min$}
	\Initial{[[0,0,...,0], ..., [0,0,...,0]], [0,0,...,0], $K$}
\EndPayloadML
\Query{\textit{value}}{}{integer $v$}
	\Let $v = min + \sum\limits_{i \in \mathit{Ids}} R[i][i] - \sum\limits_{i \in \mathit{Ids}} U[i]$
\EndQuery
\Query{\textit{localRights}}{} integer $v$
	\Let{$id = \mathit{repId}()$ \hspace{14ex} \%Id of the local replica}
	\Let{$v = R[id][id] + \sum\limits_{i \neq id}{R[i][id]} - \sum\limits_{i \neq id}{R[id][i]} - U[id]$}
\EndQuery
\Update{\textit{increment}}{integer $n$}
	\Let{$id = \mathit{repId}()$}
	\Let{$R[id][id] = R[id][id] + n$}
\EndUpdate
\Update{\textit{decrement}}{integer $n$}
	\Pre{$localRights() \geq n$}
	\Let{$id = \mathit{repId}()$}
	\Let{$U[id] = U[id] + n$}
\EndUpdate
\Update{\textit{transfer}}{integer $n$, replicaId $\mathit{to}$}: boolean $b$
	\Pre{$b = (localRights() \geq n)$}
	\Let{$\mathit{from} = \mathit{repId}()$}
	\Let{$R[\mathit{from}][\mathit{to}] := R[\mathit{from}][\mathit{to}] + n$}
\EndUpdate
\Update{\textit{merge}}{S}
	\Let $R[i][j] = max(R[i][j],S.R[i][j])$, 
	$\forall{i,j \in \mathit{Ids}}$
	\Let $U[i] = max(U[i],S.U[i])$, $\forall{i \in \mathit{Ids}}$
\EndUpdate
\end{algorithmic}
\vspace{-2ex}
\caption{\InvCounter{} for maintaining the invariant \emph{larger or equal to $K$}.}\label{fig:invcounter:spec}
\vspace{-2ex}
\end{figure}

We now detail the \InvCounter{}, a CRDT for maintaining the invariant \emph{larger or equal to $K$}. The pseudocode for \InvCounter{} is presented in Figure~\ref{fig:invcounter:spec}.

{\bf \InvCounter{} state:} 
The \InvCounter{} must maintain the necessary information to verify 
whether it is safe to locally execute operations or not.
This information consists in the rights each replica holds 
(as in the escrow transactional model \cite{escrow}).

To maintain this information, for a system with $n$ replicas,
we use two data structures.
The first, $R$ is a matrix of $n$ lines by $n$ columns with:
$R[i][i]$ recording the \emph{increments} executed at $r_i$, 
which define an equal number of rights initially assigned to $r_i$; 
$R[i][j]$ recording the rights transferred from $r_i$ to $r_j$.
The second, $U$ is a vector of $n$ lines with 
$U[i]$ recording the successful \emph{decrements} executed at $r_i$, 
which consume an equal number of rights.

For simplicity, our specification assumes every replica maintains a complete
copy of these data structures, but we later discuss how to avoid this in practice.

\begin{figure}[t]
\centering
\includegraphics[width=0.4\columnwidth]{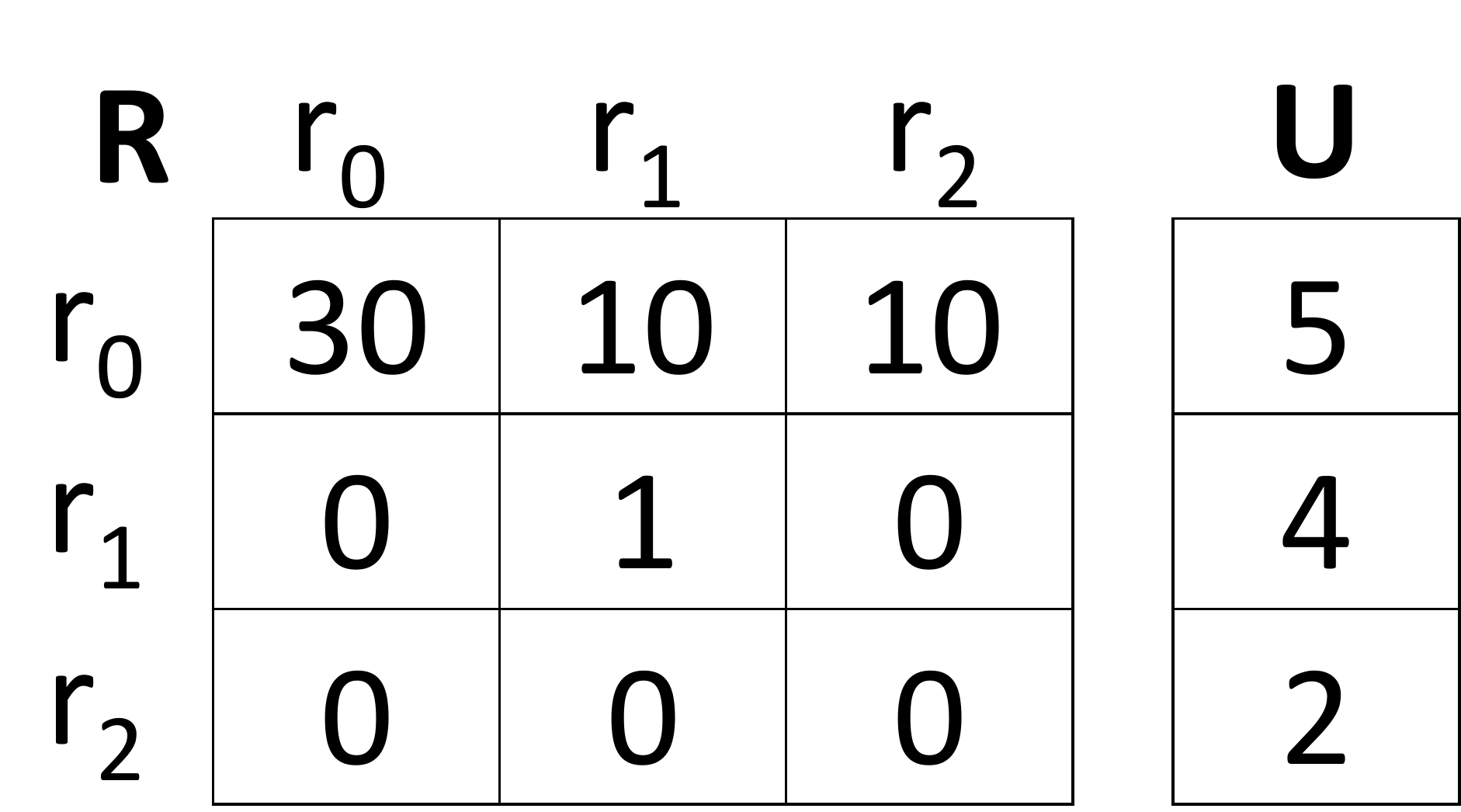}
\vspace{-2ex}
\caption{Example of the state of \InvCounter{} for maintaining the invariant \emph{larger or equal to 10}.}
\label{fig:invcounter}
\end{figure}

{\bf Operations:}
When a counter is created, we assume that the initial value of the 
counter is equal to the minimum value allowed by the invariant, $K$. 
Thus, no rights are assigned to any replica and both $R$ and $U$ are 
initialized with all entries being equal to $0$. 
To overcome the limiting assumption of the initial value being $K$, 
we can immediately execute an \emph{increment} operation in the freshly 
created \InvCounter{}.
Figure~\ref{fig:invcounter} shows an example of the state of a 
\InvCounter{} for maintaining the invariant \emph{larger or equal to 10}, 
with initial value 40. This initial value led to the creation of $30$ 
rights assigned to $r_0$ -- this value is recorded in $R[0][0]$. 

An \emph{increment} executed at $r_i$ updates the number 
of increments for $r_i$ by updating the value of $R[i][i]$.
In the example of Figure~\ref{fig:invcounter}, the value of $R[1][1]$ is $1$, 
which is the result of incrementing the counter by $1$ in $r_1$.

A \emph{decrement} executed at $r_i$ updates the number of 
decrements for $r_i$ by updating the value of $U[i]$. 
This operation can only execute if $r_i$ holds enough rights before executing
the operation. The \emph{decrement} operation fails if not enough local rights exist.
In the example of Figure~\ref{fig:invcounter}, the values of $U$ reflect the 
execution of $5$, $4$ and $2$ decrements in $r_0$, $r_1$ and $r_2$, respectively.

The rights the local replica $r_i$ holds, returned by function \emph{localRights},
are computed by:
\begin{inparaenum}[(a)]
\item adding the increments executed in the local replica, $R[i][i]$;
\item adding the rights transferred from other replicas to $r_i$, $R[j][i], \forall j \neq i$;
\item subtracting the rights transferred from $r_i$ to other replicas, $R[i][j], \forall j \neq i$; and
\item subtracting the decrements executed in $r_i$, $U[i]$.
\end{inparaenum}  
In the example of Figure~\ref{fig:invcounter}, replica $r_0$ holds 5 rights
(obtained from $30 + (0 + 0) - (10 + 10) - 5$), allowing to locally decrement the counter
by up to $5$.

The operation to retrieve the current \emph{value} of the counter consists 
of:
\begin{inparaenum}[(a)]
\item adding the minimum value, $K$;
\item adding the sum of increment operations executed at any replica, $R[i][i], \forall i$; and
\item subtracting the sum of the decrement operations executed at any replica, $U[i], \forall i$.
\end{inparaenum} 
In the example of Figure~\ref{fig:invcounter}, the current value is $30$ (obtained from $10 + (30 + 1) - (5 + 4 + 2)$).

The operation \emph{transfer} executed at replica $r_i$  
transfers rights from $r_i$ to some other replica $r_j$, by updating the
value recorded in $R[i][j]$.
This operation can only execute if enough local right exist.
In the example of Figure~\ref{fig:invcounter}, transfers of $10$ rights from 
$r_0$ to each of $r_1$ and $r_2$ are recorded in the values of $R[0][1]$ and $R[0][2]$

The \emph{merge} operation is executed during peer-to-peer synchronization,
when a replica receives the state of a remote replica. 
The local state is updated by just taking, for each entry of both data structures, 
the maximum of the local and the received value.

{\bf Correctness:}
For showing the correctness of \InvCounter{}, it is necessary to show 
that all replicas of \InvCounter{} eventually converge to the same state, i.e., that
\InvCounter{} is a correct CRDT, and 
that the execution of concurrent operations will not break the invariant. 
We now sketch an argument for why these properties are satisfied.

For showing that replicas eventually converge to the same state, 
it is necessary to prove that the specification is a monotonic semi-lattice object. 
As the elements of $R$ and $U$ are monotonically increasing (since operations never 
decrement the value of these variables), the semi-lattice properties are 
immediately satisfied -- two states, $s_0, s_1$, are related by a partial order 
relation, $s_0 \le s_1$, whenever all values of $R$ and $U$ in $s_1$ are greater 
or equal to the corresponding values in $s_0$ 
(i.e., $\forall i, j, s_0.R[i][j] \leq s_1.R[i][j] \wedge s_0.U[i] \leq s_1.U[i]$).
Furthermore, the merge of two state is the LUB, as the function just takes the maximum 

To guarantee that the invariant is not broken, it is necessary to guarantee 
that a replica does not execute an operation (\emph{decrement} or \emph{transfer}) 
without holding enough rights to do it. 
As operations execute sequentially and verify if the local replica holds enough 
rights before execution, 
it is necessary to prove that if a replica believe it has $N$ rights, 
it owns at least $N$ rights. 
The construction of the algorithms guarantees that line $i$ of $R$ and $U$ is only updated by operations 
executed at replica $r_i$. Thus, replica $r_i$ necessarily has the most recent value 
for line $i$ of both $R$ and $U$. 
As rights of replica $r_i$ are consumed by \emph{decrement} operations, recorded in $U[i]$, 
and transfer operations, recorded in $R[i][j]$, it follows immediately that replica 
$r_i$ knows of all rights it has consumed. 
Thus, when computing the local rights, the value computed locally is always conservative 
(as replica $r_i$ may not know yet of some transfer to $r_i$ executed by some other replica).
This guarantees that the invariant is not broken when operations 
execute locally in a single replica.

We wrote the specification of \InvCounter{} in TLA~\cite{tla} and 
successfully verified that the invariant holds for all the cases that the tool generated.

{\bf Extensions:}
It is possible to define a \InvCounter{} that enforces an invariant 
of the form \emph{smaller or equal to $K$} by using a similar approach, 
where rights represent the possibility of executing \emph{increment} operations 
instead of \emph{decrement} operations. 
The specification would be similar to the one presented in 
Figure~\ref{fig:invcounter:spec}, with the necessary adaptations to the 
different meaning of the rights.

A \InvCounter{} that can maintain an invariant of the form 
\emph{larger or equal to $K_0$} {\bf and} \emph{smaller or equal to $K_1$} can 
be created by combining the information of two \InvCounter{}s, 
one for each invariant, and updating both on each operation.    


{\bf Optimizations:}
The state of \InvCounter{}, as presented, has complexity $O(n^2)$. 
In practice, the impact of this is expected to be small as the number of data
centers in common deployments is typically small and each data center will 
typically hold a single logical replica.

In the cases when this is not true, we can leverage the following observations to lower the
space complexity of \InvCounters{} up to $O(n)$.
For computing the local rights, replica $r_i$ only uses the line $i$ and column $i$ of
$R$ and line $i$ of $u$. 
For computing the local value of the counter, replica $i$ 
additionally uses entries $R[i][i], \forall i$ and the remaining entries of $U$.
This leads to a space complexity of $4.n$ for storage, which compares with $2.n$ as the minimal
complexity of a state-based counter \cite{Burckhardt14Replicated}.

In this case, when synchronizing, a replica only needs to send the information both replicas
store. Thus, a replica $r_i$ would send to $r_j$ only $R[i][i], \forall i$, 
$R[i][j]$ and $U$\footnote{Replicas $r_i$ and $r_j$ also share $R[j][i]$, but 
as this value is only updated at $r_j$, it is not necessary to send it.}, 
lowering the space complexity for messages to $2.n$.

When this optimization is not in place, and 
every replica maintains the complete data structure, we can still lower
the communication costs by propagating the information
epidemically. This means that it is not necessary for every replica to
communicate directly with every other replica.
In particular, we can allow for the communication to be reactive instead
of proactive: a replica $r_i$ only needs to communicate directly with $r_j$
when it transfers rights to $r_j$ (e.g., upon request in order to
execute an operation) so that $r_j$
knows about the new rights. Note that the lack of communication does
not affect the correctness regarding the invariant violation, as each replica 
always has a conservative view on its available rights. 

\section{Middleware for Enfocing Numeric Invariants}\label{sec:deploy:riak}

We now present two middleware designs for extending Riak database with numeric
invariants, using the \InvCounter{}.
The proposed designs can be applied to any database that provides the following
two properties, essential for \InvCounter{} to work properly.
First, each replica needs to execute operations referring to each counter in a serializable way, i.e., as if they had been executed in sequence. This does not, however, preclude concurrency: operations for different counters are not constrained by this requirement, and even within the same counter there are protocols that allow for some concurrency while maintaining the illusion of a serializable execution.
This serialization is necessary to guarantee that two concurrent operations do not use the same
rights.
Second, the replication model must ensure no lost updates, i.e., updates executed
concurrently in different replicas must be merged using the CRDT merge function.
This is necessary for the CRDT to work properly.

Before presenting the middleware designs, we present an overview of the 
functionalities of Riak that are relevant for the deployment of \InvCounters{}.


\subsection{Overview of Riak 2.0}

Riak 2.0 is a key/value database inspired in Dynamo~\cite{dynamo}.
It support geo-replication in its Enterprise Edition, where each DC 
maintains a full replica of the database.
Riak provides an API supporting a read (\emph{get}) and write (\emph{put})
interface, where a
write associates a new value with a key, and a read
returns the value(s) associated with the key.

By default, writes on a key can proceed concurrently, with the system
maintaining the multiple concurrent versions and exposing them to clients
in subsequent read operations.
Additionally, Riak includes native support for storing CRDTs, dubbed Riak data types,
where concurrent writes are automatically merged. 

Riak keys can be marked as strongly consistent. For these keys, 
Riak uses a conditional writing mode where a write fails if a 
concurrent write has been executed.
These key are not geo-replicated (each DC has its local view
of the data) and they cannot store a Riak data type object.

\subsection{Alternative 1: Client-based Middleware}\label{sec:deploy:cltlib}

\begin{figure}
\includegraphics[width=0.9\columnwidth]{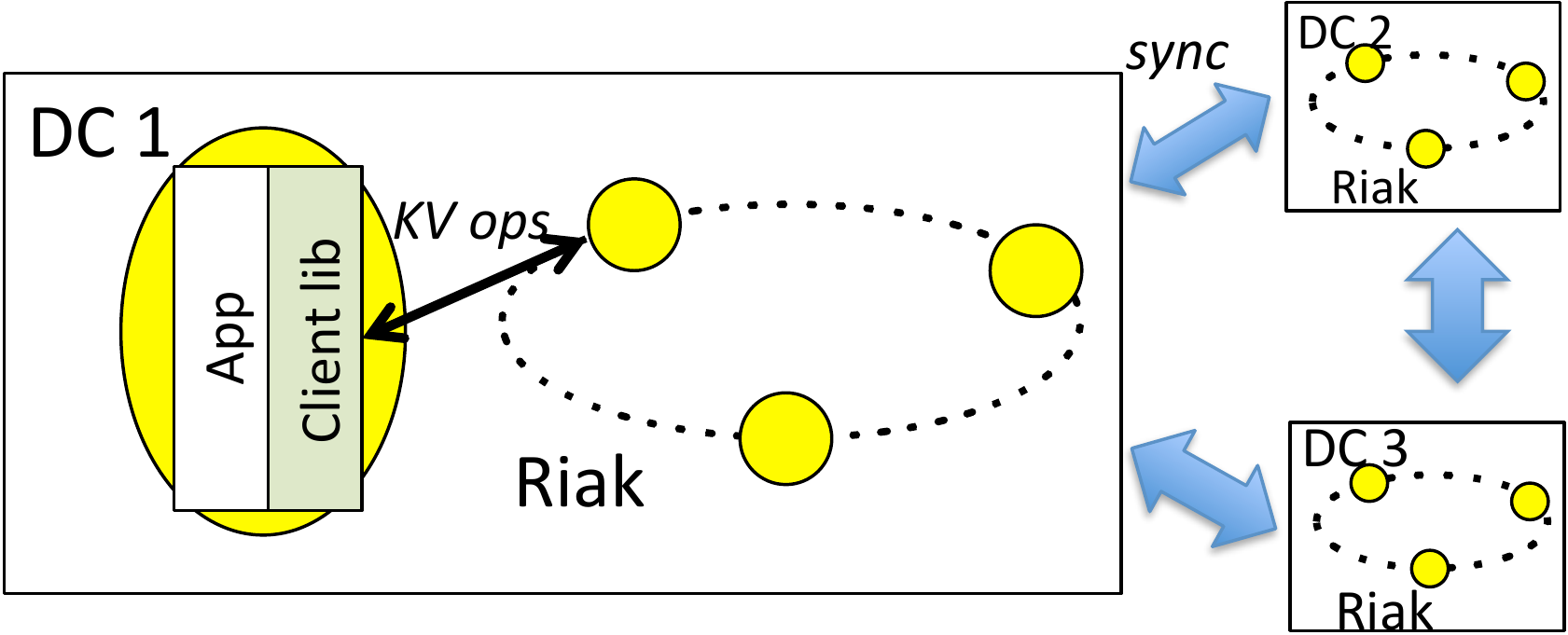}
\vspace{-2ex}
\caption{Client-based middleware for deploying \InvCounters{}.}\label{fig:middle:library}
\end{figure}

Our first design, depicted in Figure \ref{fig:middle:library}, is based on a client-side middleware.
Supporting operations on \InvCounters{} is fairly simple, given the functionality provided by Riak.

The state of a \InvCounter{} is stored as an opaque object in the Riak database, which
is marked as strongly consistent.
Rights for executing operations in a \InvCounter{} are associated with each DC, i.e., 
each DC is considered as a single replica for a \InvCounter{}. 
An increment (resp.\ decrement) executes in the client library by first reading 
the current value of the counter (executing a \emph{get} operation in Riak), 
then executing the increment (resp.\ decrement) operation in the \InvCounter{} 
and writing the new value of the counter back into the database. 
If the operation in the \InvCounter{} fails, the client can try to obtain additional 
rights by requesting the execution of a \emph{transfer} operation from another DC. 
If the operation in the CRDT succeeds but the conditional write fails, the operation 
must be re-executed until it succeeds. 

Given that \InvCounters{} are marked as strongly consistent, updates are serialized 
in each DC through the conditional writing mechanism.
Concurrent updates to the same \InvCounter{} can only 
appear due to geo-replication. 
If this is the case, then concurrent versions can be merged by the client 
library when reading the counter.

For propagating the updated values across DC, we were not able to reuse 
the geo-replication mechanism from Riak, since it does not support multi-data 
center replication for objects that use strong consistency. 
As such, we had to implement a custom synchronization mechanism for 
\InvCounters{}.
This custom synchronization forwards modified counters to other DCs periodically.
A DC receiving a remote version of a counter, merges the received version with 
the local version. 

\begin{figure}
\includegraphics[width=0.9\columnwidth]{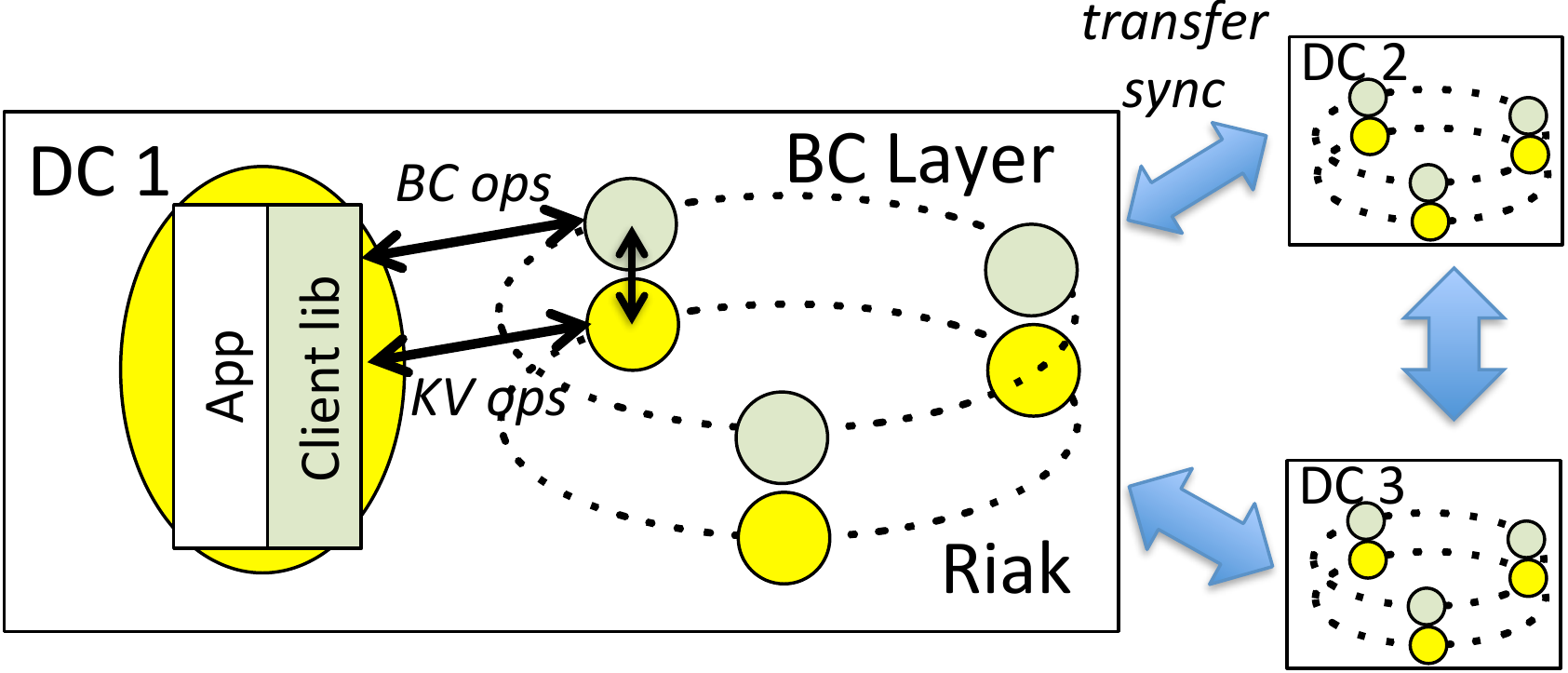}
\vspace{-2ex}
\caption{Server-based middleware for deploying \InvCounters{}.}\label{fig:middle-server}
\vspace{-2ex}
\end{figure}

\subsection{Alternative 2: Server-based Middleware}\label{sec:deploy:srvmid}

The client-based middleware has an important limitation, as pointed out by the evaluation in Section~\ref{sec:eval}: 
the conditional writing mechanism for serializing operation execution works well under low load, 
but leads to an increased number of failed writes when the load increases. 
To address this issue, we propose a server-based middleware design that serializes
all operations executed in each DC for each counter.

The server-based middleware is built using a DHT communication substrate 
(riak\_core \cite{riakcore} in our prototype) running side by side with each
node of the Riak database. The key feature 
that is employed is the ability to {\em lookup} the DHT node that is 
responsible for a given key. 
This primitive is used to route all requests for a given key to the same node,
which serializes their execution.
For operations on regular objects, the client library calls Riak directly 
(without contacting DHT nodes). 

When an application wants to execute an operation in a counter, 
the operation is sent to the DHT node responsible for that counter. 
The DHT node executes the operation by reading the counter from Riak, executing
the operation and writing back the new value. \InvCounters{} are
marked as strongly consistent, with writes being executed using conditional write.
In the normal case, when there are no reconfigurations, the conditional write 
will succeed, since a single DHT node is responsible for any given key 
and executes all operations for each counter in sequence.

In contrast, when a new nodes enters the DHT or some node fails, 
the DHT is automatically reconfigured and it
becomes possible that two nodes concurrently
process two operations for the same key.
In this case, only the first write will succeed, since the following concurrent
writes will fail due to the conditional write mechanism.
This guarantees the correctness of the system, by serializing all
updates.

Since Riak does not geo-replicate keys marked as strongly consistent,
our middleware had to include a mechanism for propagating updates to \InvCounters{} 
to other DCs. 
To this end, each DHT node periodically propagates its updated entries to 
the corresponding DHT nodes in other DCs. 
With this approach, each value that is sent can include the effects of 
a sequence of operations, thus reducing the communication overhead.
As in the previous version, when a \InvCounter{} is received in a DC from another DC,
it is merged with the local replica using the CRDT merge function.
For other objects, we rely on normal built-in Riak multi-data center replication.

{\bf Optimizations:} 
Our prototype includes a number of optimization to improve its efficiency.
The first optimization is to cache \InvCounters{} at the middleware layer. 
This allows us to reduce the number of Riak operations necessary for processing 
each update on a \InvCounter{} from two to one -- only the write is necessary.

Under high contention in a \InvCounter{}, the design described so far is not very efficient, 
since an operation must complete before the next operation starts being processed. 
In particular, since processing an update requires writing the modified \InvCounter{}
back to the Riak database, which involves contacting remote nodes, each operation can 
take a few milliseconds to complete. 
To improve throughput, while a remote write to Riak is taking place, the operations 
that are received are executed in the local copy of the \InvCounter{}. 
If the counter cannot be incremented or decremented, the result is immediately returned to the client. 
Otherwise, no result is immediately returned and the operation becomes pending. 
When the previous write to the Riak database completes, the local version of 
the \InvCounter{}, which absorbed the modifications of all pending operations, 
is written in the Riak database.
If this second conditional write succeeds, all pending operations complete by 
returning success to the clients. 
Otherwise, clients are notified of the failure.

\subsection{Transferring Rights}

For executing an operation that may violate an invariant, a replica needs to own 
enough rights.
Given that it is impossible to anticipate the rights needed at each replica, it is
necessary to redistribute rights among replicas.

In our middleware designs, replicas proactively exchange 
rights in the background.
A replica that has fewer rights than a given threshold periodically asks
additional rights from replicas that have more rights
(as reflected in the local replica of the \InvCounter{}).
The number of rights requested is half of the difference between the rights 
of the remote and the local replicas.
A replica receiving an asynchronous transfer request never accepts to transfer 
more than half of the available rights.
This strategy provisions replicas with rights without impairing the latency 
during operation execution.

Nonetheless, it may happen that an operation does not succeed because it has insufficient
local rights during execution.
In this situation, the programmer can choose to get the rights from a remote replica
or abort the operation.
Programmatically the decision is made through the \emph{flag} parameter 
in the \emph{decrement} and \emph{increment} operations, as presented in Figure~\ref{fig:API}.

To execute a transfer, replica $r_i$ checks the local state
to choose the best candidate replica to request rights from (e.g., the remote replica holding more rights), 
$r_j$, and sends a transfer request and a flag saying whether it is a synchronous or an asynchronous request.
Upon receiving the request, the remote replica $r_j$ checks if it can satisfy the request
and if so it executes a local \emph{transfer} operation to move
the rights from $r_j$ to $r_i$. 
If the request was asynchronous the replication 
mechanism will asynchronously propagate the update to the requester, 
otherwise $r_j$ stores the transfer locally and replies to $r_i$ 
immediately with the the new state of the counter.

Replying to every transfer request may lead to a request being satisfied more 
than once, either because a request message was lost and replayed or because the 
requester sent the request more than once (possibly to multiple replicas).
To avoid this situation, $r_i$ sends in the request to $r_j$ the number 
of rights transferred from $r_j$ to $r_i$ ($R[j][i]$).
The receiver ignores a request if it has already transferred more rights.

A property of the way \emph{transfer} is implemented is that it does not 
require any strong synchronization between the replica asking for rights and 
the one providing the rights. 
Thus, the request for a transfer and synchronization of the information 
about transferred values can be done completely asynchronously, 
which simplifies the system design. 

\subsection{Fault-tolerance}

We now analyze how our middleware designs provide fault-tolerance
building on the fault-tolerance of the underlying cloud
database. 
We start by noting that for the \InvCounters{}, each DC acts as a 
\InvCounter{} replica.

A DC is assumed to have sufficient internal redundancy to never lose its state.
In Riak, the level of fault-tolerance in each DC can be controlled by 
changing the size of the quorums used to store data.
Thus, an update to an \InvCounter{} executed in a DC is never lost 
unless the DC fails forever.

A failure in a node in the DC may cause the DHT used in our server-based
middleware to reconfigure. As we explained before, this does not affect
correctness as we rely on conditional writes to guarantee that operations
of each counter are serialized in each DC.

During a network partition, rights can be used in both sides of the partition -- the
only restriction is that it is impossible to transfer rights between
any two partitioned DCs.
If an entire DC becomes unavailable,
the rights owned by the unreachable DC become temporarily
unavailable.
If a DC fails permanently, as the \InvCounter{} records the rights owned 
by every replica, it is possible to recover the rights that were 
owned by the failed DC.

\section{Evaluation}\label{sec:eval}

We implemented both middleware designs for extending Riak with numeric invariants and evaluated experimentally the prototypes.
This evaluation tries to address the following main questions.
\begin{inparaenum}[(i)]
\item How much overhead is introduced by our designs?
\item What is the performance penalty when the bounds are close
to being exceeded? 
\item How does the performance vary with the level of contention for
the same counter?
\end{inparaenum}

In our designs, operations on \InvCounters{} are handled by our middleware. All other operations
are directly executed in the Riak database.
For this reason, our evaluation focus on the performance of \InvCounters{}, using micro-benchmarks 
to test different properties of the system.

\subsection{Configurations}

In the experiments, we compare the client-based middleware, \emph{BCclt}, and
sever-based middleware, \emph{BCsrv}, with the following 
configurations.

\emph{Weakly Consistent Counters (Weak).} 
This configuration uses Riak's native counters operating under weak consistency. 
Before issuing a decrement, a client reads the current counter value and issue 
a decrement only if the value is positive.


\emph{Strongly Consistent Counters (Strong).} 
This configuration uses Riak's native strong consistency, with
the Riak database running in a single DC, which 
receives requests from clients in the local and remote DCs.
As Riak strong consistency cannot be used with Riak data types, the
value of the counter is stored as an opaque object for Riak.
A counter is updated by reading its value, updating its state if the value is
positive, and writing back the new state (using a conditional write).


\subsection{Experimental Setup}

Our experiments comprised 3 Amazon EC2 DCs distributed across the globe.
The latency between each DC is shown in Table \ref{table:latency_table}.
In each DC, we use three m1.large machines with 7.5GB of memory for running 
the database servers and server-based middleware and three m1.large machines
for running the clients.

\begin{table}[h]
\center
\begin{tabular}{llll}
RTT (ms) & US-E & US-W & EU \\
US-East    & -        & 80   & 96        \\
US-West      & 83       & -      & 163     \\
EU      & 93     & 161    & -        
\end{tabular}
\caption{RTT Latency between Data Centers in Amazon EC2.}
\label{table:latency_table}
\end{table}

For \emph{Weak}, we used Riak 2.0 Enterprise Edition (EE), with support
for geo-replication. 
For other configurations we used Riak 2.0 Community Edition (CE), with support for
strong consistency. 
Both version share the same code, except for the support for strong consistency and
geo-replication, which is only available in the enterprise edition.

In \emph{Strong}, data is stored in the US-East DC, which is the location that
minimizes the latency for remote clients.
In the remaining configurations, data is fully geo-replicated in all DCs, with
clients accessing the replicas in the local DC.
Riak operations use a quorum of 3 replicas for writes and 1 replica for reads.

\subsection{Single Counter}
\label{sec:single_counter}

The objective of this experiment is to evaluate the performance of the middleware 
designs in contention scenarios.
In this case, we use a single counter initialized to a value that is large enough to
never break the invariant ($10^9$). Clients execute 20\% of increments and 80\% of decrements 
in a closed loop with a think time of 100 ms. Each experiment runs for two minutes 
after the initialization of the database.
The load is controlled by tuning the number of clients running in each 
experiment -- clients are always evenly distributed among the the client machines.
 
\begin{figure}[t]\centering
\includegraphics[width=0.8\columnwidth]{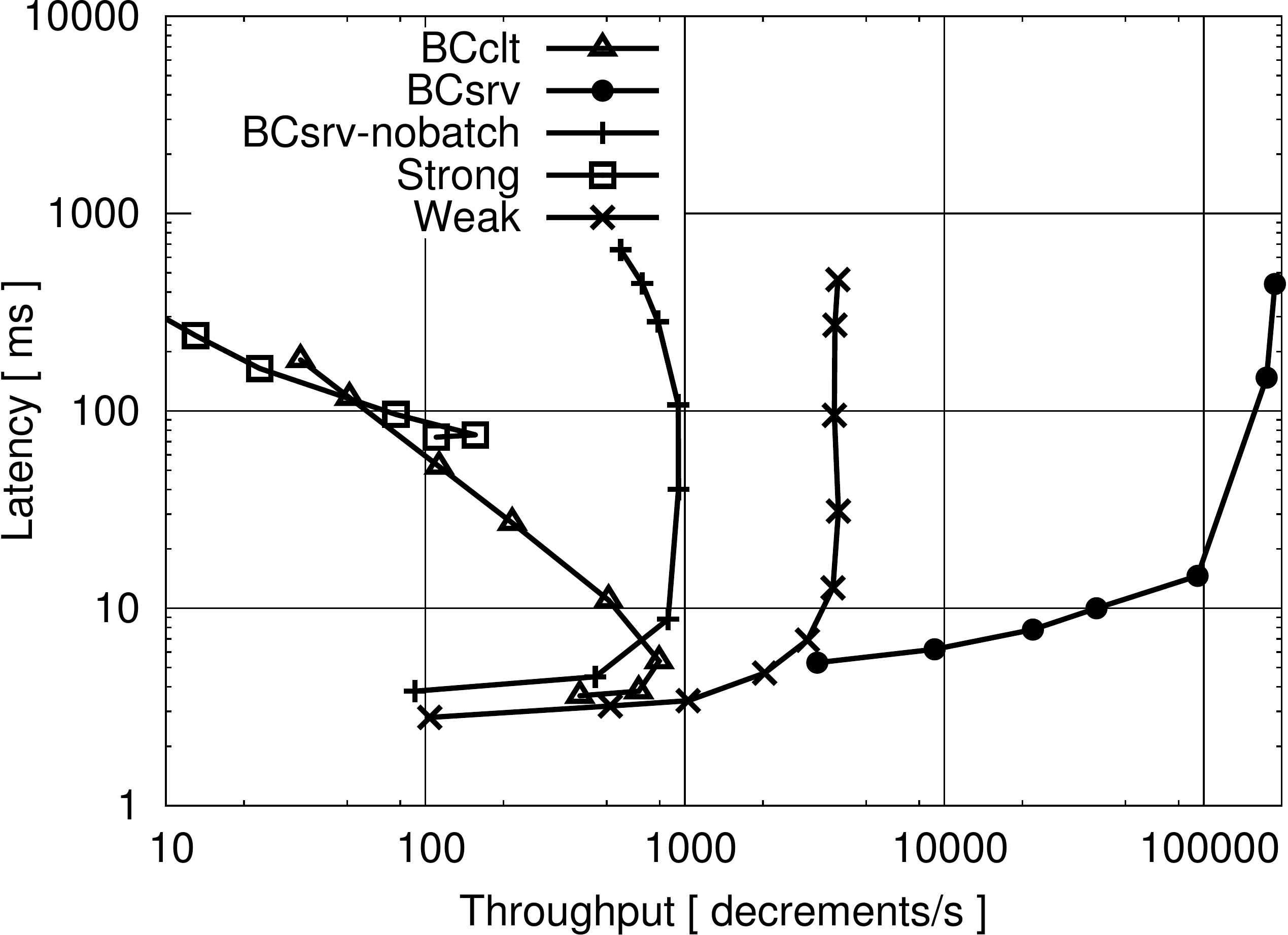}
\vspace{-1ex}
\caption{Throughput vs. latency with a single counter.}
\label{fig:single:throughput}
\end{figure}


\begin{figure*}[t]
\centering
\begin{minipage}{.29\textwidth}
\includegraphics[width=\columnwidth]{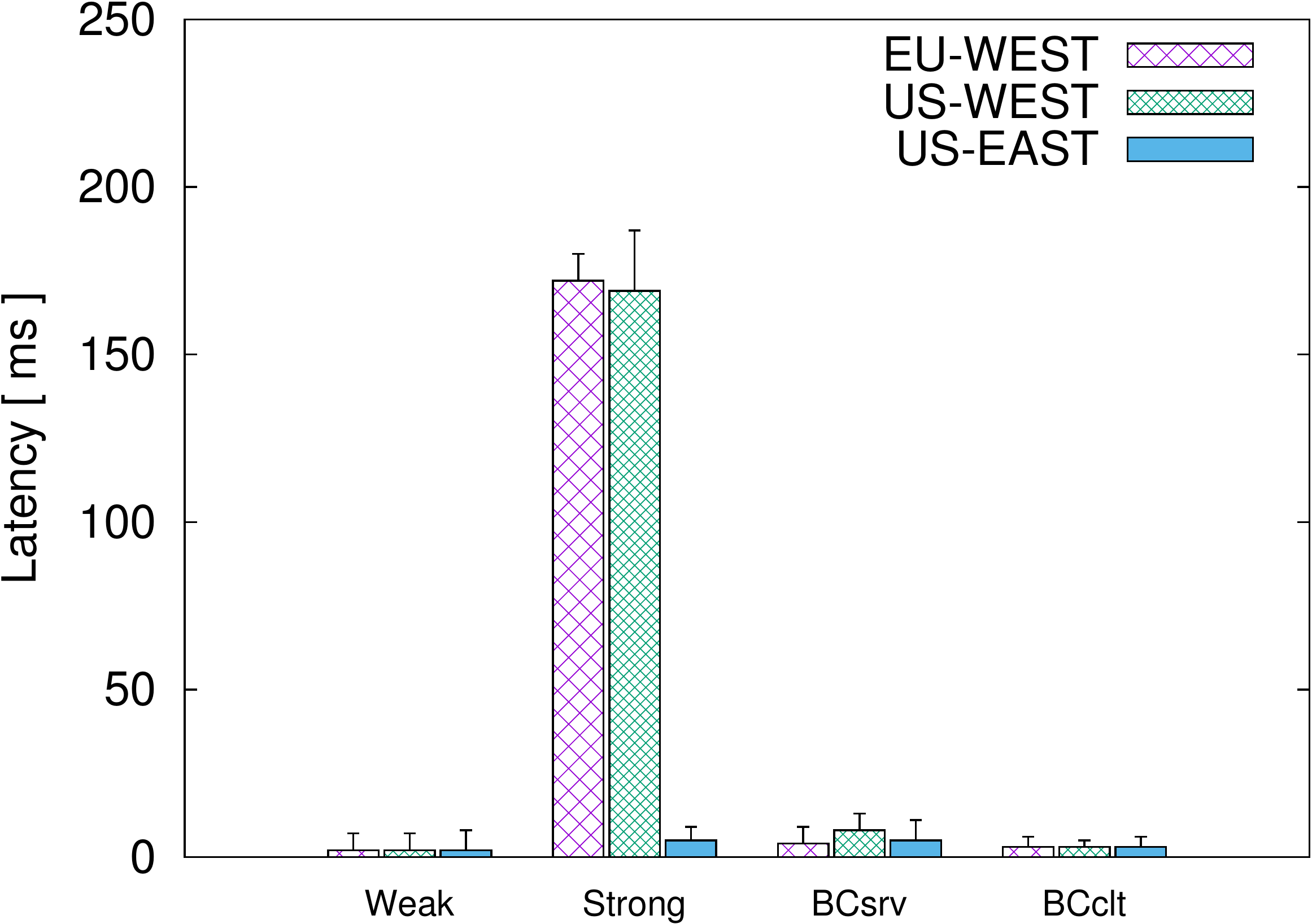}
\caption{Median latency with a single counter, per region of clients (the line is the value for the $99^{th}$ percentile).}
\label{fig:single:latencybar}
\end{minipage}%
\hspace{0.04\textwidth}
\begin{minipage}{.29\textwidth}
\includegraphics[width=\columnwidth]{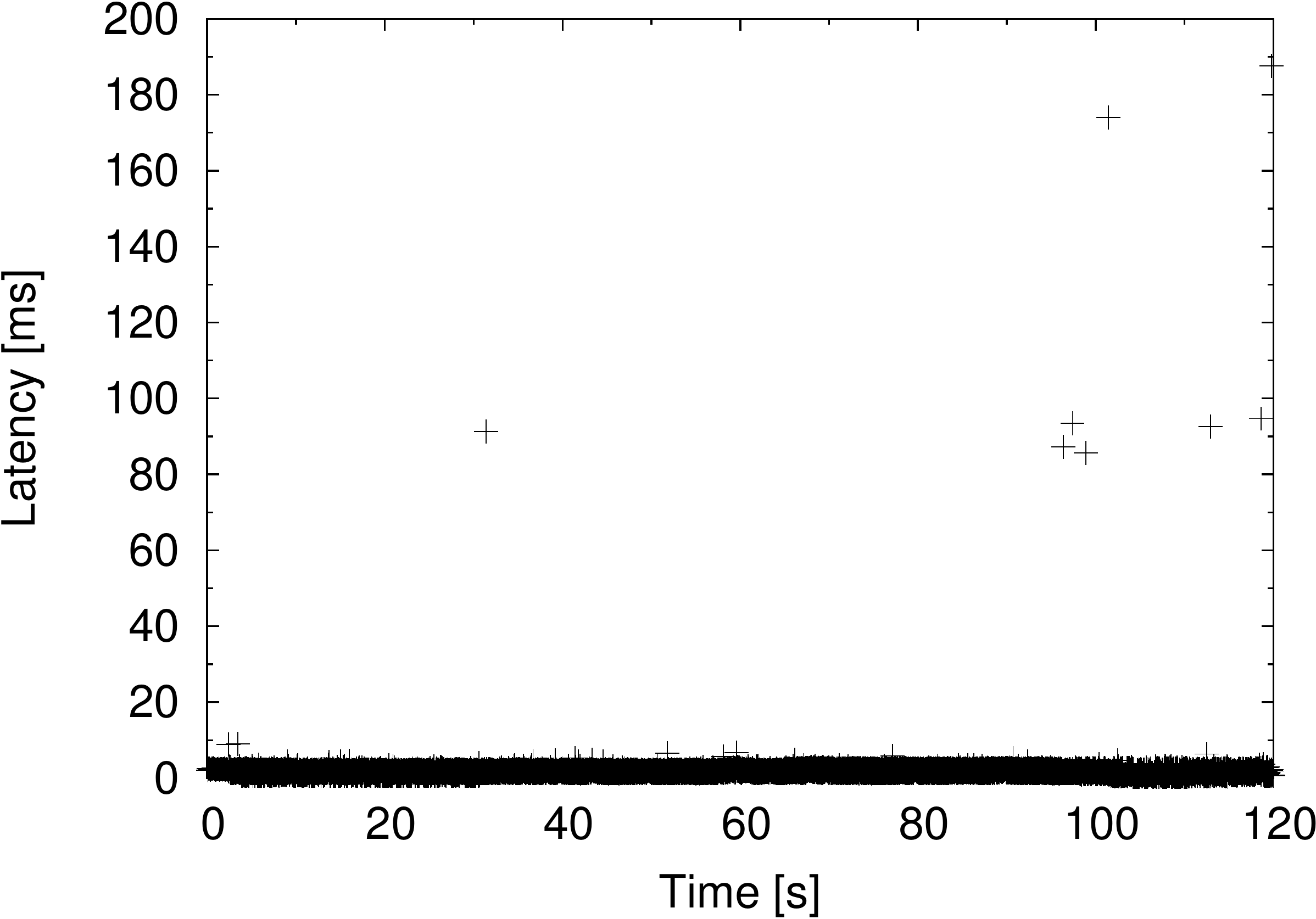}
\caption{Latency of each operation over time for BCsrv.}
\label{fig:single:latOvertime}
\end{minipage}%
\hspace{0.04\textwidth}
\begin{minipage}{.29\textwidth}
\includegraphics[width=\columnwidth]{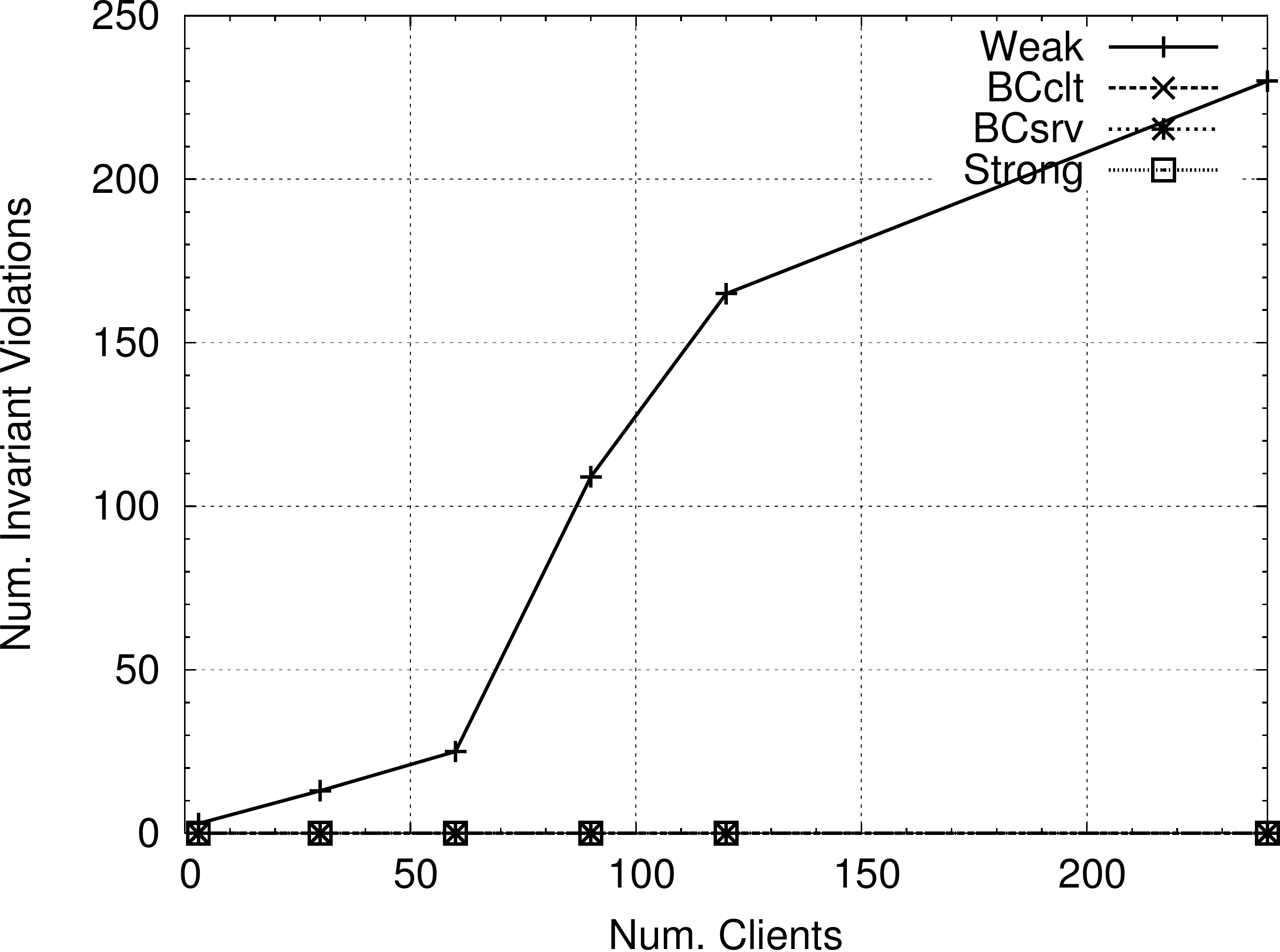}
\caption{Decrements executed in excess, violating invariant.}
\label{fig:single:excess}
\end{minipage}%
\end{figure*}

\paragraph{Throughput vs. latency}
Figure~\ref{fig:single:throughput} presents the variation of the throughput vs.\ latency values as
more operations are injected in the system.
For the throughput values we consider only the operations that have succeeded, while for
the latency we consider the average of all (succeeded or failed) operations. (This only affects the results for \emph{Strong}.)

The results of \emph{BCclt} and \emph{Strong} present a similar trend, which
is that the
throughput quickly starts degrading with the increase in the load.
By analyzing the results of the operations, we found out that this is explained
by the fact that the percentage of operations that fail increase very quickly with
the number of clients. This is because concurrent updates fail due to the conditional write 
mechanism -- e.g., for \emph{Strong}, 50\% of operations 
fail with 100 clients and 90\% with 200 clients.
The $3\times$ higher throughput in \emph{BCclt} is explained by the fact that 
clients execute operations in their local DC, while in \emph{Strong} all operations
are sent to a single DC. 
The higher average latency in \emph{Strong} is explained by the latency of
operations from remote clients. 
This explains why we chose to report the latency 
of all operations, including failed ones: since most of remote operations fail, considering
only operations that succeed would lead to latency values close to those of
\emph{BCclt}.

The throughput of \emph{Weak} is much larger and it does not degrade with
the increase of the load -- when it reaches its maximum throughput, increasing
the load just leads to an increase in latency. 
Our server-based middleware, \emph{BCsrv}, has an even higher throughput with slightly
higher latency. 
The higher latency is expected, as the middleware introduces communication overhead.
The higher throughput is due to the batching mechanism introduces in \emph{BCsrv},
which batches a sequence of updates into a single Riak write,
thus leading to a constant rate of Riak operations.
To prove this hypothesis, we have run the same experiment, turning off the batching and writing
every update in Riak - results
are presented as \emph{BCsrv-nobatch}.
In this case, we can observe that the throughput is much lower than \emph{Weak}, but unlike
\emph{BCclt}, the throughput does not degrade with the load - the reason for this is that
the middleware serializes updates and Riak still sees a constant rate of writes. 
The same approach for batching multiple operations into a single Riak write could
be used with other configurations, such as \emph{Weak},
to improve their scalability.

\paragraph{Latency under low load}
Figure~\ref{fig:single:latencybar} presents the median latency experienced by clients in different
regions when load is low (with 15 threads in each client machine).
As expected, the results show that for \emph{Strong}, remote clients experience high latency 
for operation execution, while local clients are fast.
The latency for all the other configurations is very low, with \emph{BCsrv} introducing a
slight overhead (of about 2 ms), due to additional communication steps for processing the request. 
If \InvCounters{} were added to the Riak database, this overhead could be eliminated.

\paragraph{Effects of exhausting rights}
In this experiment we evaluate the behavior of our middleware when the
value of the counter approaches the limit.
To this end, we run the experiment with \emph{BCsrv} and 5 clients executing $100\%$ decrements, 
initializing the counter with the value 6000 and running an experiment  
until the rights are all consumed. 

Figure \ref{fig:single:latOvertime} shows that most operations take low latency, with a few 
peaks of high latency whenever a replica needs to obtain additional rights.
The number of peaks is small because most of the time the proactive mechanism 
for exchanging rights is able to provision a replica with enough rights before 
all rights are used. 

These peaks can occur at any time during the experiment, but are more frequent when
the resources are low and replicas exchange rights more often -- close to 
the end of the experiment. 
After all rights are consumed, the latency remains low because a replica 
does not ask for rights from replicas that are expected to have no rights (according
to the local copy of the \InvCounter{}). Thus, when all rights are consumed,
operations fail locally. 


\paragraph{Invariant Preservation}
To evaluate the severity of the risk of invariant violation,
we computed how may decrements in excess were executed with success in the different solutions. 
The counter is initialized with the value of $6,000$ and a variable number of clients 
execute decrement operations with a think time 100 ms.
Figure~\ref{fig:single:excess} shows that the invariant was only broken in \emph{Weak}, as expected.
The figure shows that an increase in the number of clients directly impacts the severity of
the invariant violation. This is because
in \emph{Weak} the client reads a counter, checks if its value is greater than 
the limit and decrements it.
Since this is not an atomic operation, the value of the counter can be different between 
the read and the update, and that difference is directly affected by the number of concurrent updates,
which leads to more invariant violations.


\begin{figure}[t]\centering
\includegraphics[width=0.8\columnwidth]{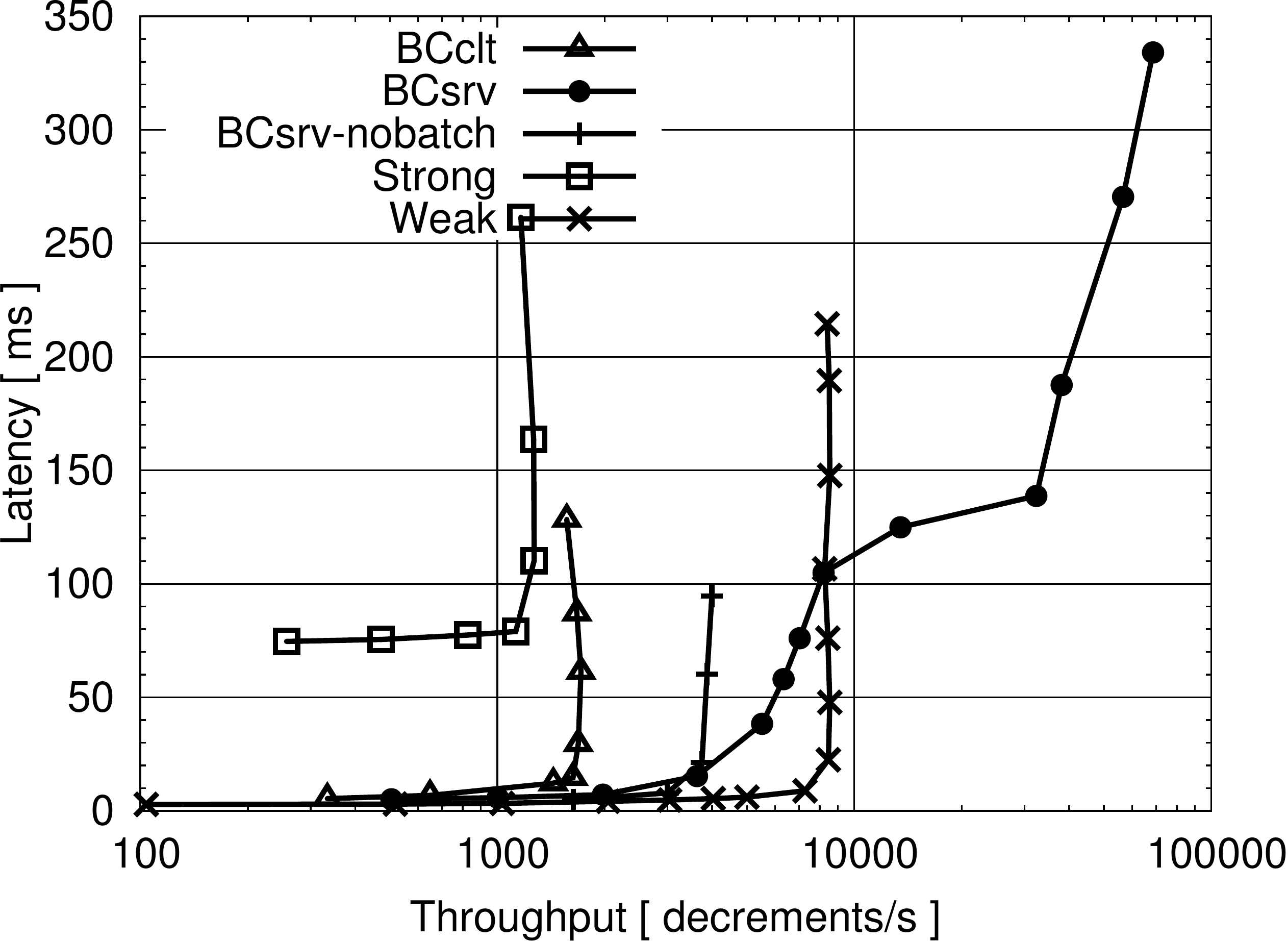}
\vspace{-1ex}
\caption{Throughput vs. latency with multiple counters.}
\label{fig:multiple:throughput}
\end{figure}

\subsection{Multiple Counters}

To evaluate how the system behaves in the common case where clients access to
multiple counters, we ran the same experiment of 
Section \ref{sec:single_counter} with 100 counters.
For each operation, a client selects the counter to update randomly
with uniform distribution. 
Increasing the number of counters reduces the contention in each key and 
contributes to balance the load among nodes.

The results presented in Figure~\ref{fig:multiple:throughput} show that 
both \emph{BCclt} and \emph{Strong} now scale to a larger throughput (when
compared with the results with a single key). 
The reason for this is that by increasing the number of counters, the 
number of concurrent writes to the same key is lower, leading
to a smaller number of failed operations.
This in turn increases with the load, as expected. 
Additionally, when the maximum throughput is reached, the
latency degrades but the throughput remains almost constant. 
The higher average latency in \emph{Strong} is explained by the fact that remote
operations have high latency, as shown before.

The \emph{Weak} configuration scales up to a much larger value (9K decrements/s 
compared with 3K decrements/s for a single counter).
As each Riak node includes multiple virtual nodes, when using multiple counters
the load is balanced among them - enabling multi-core 
capabilities to process multiple requests in parallel (whereas with a single node, a single virtual node is used, resulting in  
requests being processed sequentially).


The results show that \emph{BCsrv} has a low latency close to \emph{Weak}'s
as long the number of writes can be handled by Riak's
strong consistency mode in a timely manner. 
In contrast with the experiment with a single counter, Riak's capacity is shared among all the keys, each contributing with writes to Riak.
Therefore, as the load increases, writing batches to Riak will take longer to complete and contribute to accumulate latency sooner than in the single key case.
Nevertheless, batching still allows multiple client requests to be processed per each Riak operation, leading to a better throughput.
The maximum throughput even surpasses the results for the \emph{Weak} configuration.
The results for \emph{BCsrv-nobatch}, where each individual update is written 
using one Riak operation, can be seen as the worst case of our middleware, 
in which the batching had no effect. Still, since all \emph{BCsrv} operations are local to a given DC and access only a quorum of Riak nodes, one
can expect that increasing the local cluster's capacity should have a positive effect both on latency and throughput.

\section{Discussion}~\label{sec:extension}

In this section we discuss how to extend our approach,
to support other could databases and additional invariants.

\subsection{Supporting Other Cloud Databases}

Although our middleware designs run on top
of the Riak database, it would be immediate to implement a similar
prototype running on top of any database that 
provides conditional writes, such as DynamoDB~\cite{dynamo}. 
Given that we had to implement the geo-replication in the middleware, we do 
not even require native support for geo-replication.

Alternatively, if the database provides a way to serialize 
all operations to a given key, it would be easy to adapt the current design. 
We note that this could be done in two different ways: either the 
cloud database already supports these strong semantics, 
in which case there is no need to add any further logic, 
or the DHT has a way to ensure that messages routed to a given 
key are delivered in sequence, in which case the DHT can keep 
track of the latest operation issued to the database.

\subsection{Supporting Other Invariants}

Some applications might require that a
a counter is involved in more than one numeric invariant, and also that some
invariants refer to multiple counters -- e.g., we may want to have $x \geq 0 \wedge y \geq 0 \wedge x + y \geq K$.
To address this, the invariant $x + y \geq K$ can be maintained by a \InvCounter{} that represents the value of $x + y$.
In this case, when updating the value of $x$ (resp.\ $y$), it is necessary to update 
both the \InvCounter{} for $x$ (resp.\ $y$) and for $x + y$, with the operation succeeding 
if both execute with success. 
For maintaining such invariants, this needs to be done atomically but not in isolation. In other words,
either both \InvCounters{} are updated or none, however, it is safe for an application to observe 
a state where only one of the \InvCounters{} has been updated.

Without considering failures, this allows for a simple implementation where, if one
\InvCounter{} operation fails, the operation in the other \InvCounter{} is  
compensated \cite{sagas} by executing the inverse operation.
When considering failures, it is necessary to include some transactional mechanism 
for guaranteeing that either both updates execute or none -- recently, eventually consistent
cloud databases started to support such features \cite{cops,eiger}.

A number of other invariants, such as referential integrity and 
key constraints, can be encoded as numeric invariants,
as discussed by Barbará-Milla and Garcia-Molina \cite{demarcation}.
Those approaches could be adapted for using \InvCounters{}.

\section{Related work}\label{sec:related}
\label{sec:related-work}

Many cloud databases supporting geo-replication have been developed in recent years.
Several of them~\cite{dynamo,cops,eiger,chainreaction,cassandra,riak,dynamoDB} offer variants of eventual/weak consistency where operations return immediately once executed in a single data center. Such approach is favored for the low latency it can achieve when it selects a data center close the end-user (among several scattered across the world). Each variant addresses particular requirements, such as: reading a causally consistent view of the database \cite{cops,chainreaction}; writing a set of updates atomically\cite{eiger}; or, supporting application-specific or type-specific reconciliation with no lost updates\cite{dynamo,cops,walter,dynamoDB,riak}.
Our work focuses on the complementary requirement of having counters that enforce a global numeric invariant.

For some applications eventual consistency needs to be complemented or replaced with strong consistency to ensure correctness. Spanner \cite{spanner} provides strong consistency for the whole database, at the cost of high coordination overhead for all updates. Transaction chains \cite{transactionchains} is an alternative that offers transaction serializability with latency proportional to the latency to the first replica accessed. 

Often, only specific operations require strong consistency. Walter \cite{walter} and RedBlue consistency in Gemini \cite{redblue} can mix eventual and strong consistency (snapshot isolation in Walter) to allow eventually consistent operations to be fast. PNUTS \cite{pnuts}, DynamoDB \cite{dynamoDB} and Riak \cite{riak} also combine weak consistency with per-object strong consistency, by relying on conditional writes that fail if concurrent ones existed.
Megastore \cite{megastore} offers strong consistency inside a partition
and weak consistency across partitions. In contrast, our work extends eventual consistency with numeric invariants. This allows, for the specific case of applications that require numeric invariants to be preserved, their correctness to be met while still allowing most operations to execute in a single replica.

Bailis et al.~\cite{coordinationavoiding} examine which operations in database systems require coordination for meeting invariants. 
We provide a low cost solution for operations that may break numeric invariants, which require coordination under their analysis. This is possible because we secure the necessary rights prior to executing the operations, and this way move coordination outside the critical path of operation execution.

Escrow transactions \cite{escrow}, initially
proposed for increasing concurrency of transactions in single databases, have also been
used for supporting disconnected operation in mobile computing environments 
either relying on centralized \cite{mobisnap,Walborn95Semantics} or peer-to-peer \cite{exo-leasing} 
protocols for escrow distribution. 
The demarcation protocol \cite{demarcation} enforces invariants across multiple objects,
located in different nodes.
The underlying protocols are similar to escrow-based ones, with peer-to-peer interaction. 
MDCC~\cite{mdcc} has recently proposed a variant of this protocol
for enforcing data invariants in quorum systems.

Our work combines convergent data-types \cite{crdts} with ideas from 
these systems to provide a decentralized approach with replicated data that offers
both automatic convergence and invariant-preservation with no central authority.
Additionally, we describe, implement and evaluate how such solution can be integrated into
existing eventually consistent cloud databases.


Warranties \cite{warranties} provide time-limited assertions over the state of the database
and have been used for improving latency of read operations in cloud databases.
The goal of warranties is to support linearizability efficiently, whereas ours is
to permit concurrent updates while enforcing invariants.

\section{Conclusion}\label{sec:conclusions}

This paper proposed two middleware designs for extending eventually consistent
cloud databases with the ability to enforce numeric invariants.
Our designs allow most operations to complete within a single DC by moving
the necessary coordination outside of the critical
path of operation execution, thus combining 
the benefits of eventual consistency -- low latency, high availability --
with those of strong consistency -- enforcing global invariants.
The evaluation of our prototypes shows that our client-based middleware 
does not scale when contention is high, 
but our server-based middleware, featuring a cache and a write batching mechanism,
scales even better than the Riak's native 
weak consistency mechanism where invariants can be compromised.


\section*{Acknowledgments}

This research is supported in part 
    by EU FP7 SyncFree project (609551),
    FCT/MCT SFRH/BD/87540/2012, PTDC/ EEI-SCR/ 1837/ 2012 and PEst-OE/ EEI/ UI0527/ 2014. The research of Rodrigo\ Rodrigues is supported by the European Research Council under an ERC Starting Grant.

{\bibliographystyle{acm}


\end{document}